\documentclass[12pt]{article}
\usepackage[T1]{fontenc}
\usepackage{graphicx}
\newcommand{\interlinia}{}

\title{Theory of the dielectric susceptibility
of liquid crystals with bent-core molecules}

\author{A. Kapanowski \\
{\em Institute of Physics, Jagiellonian University,}\\
{\em ulica Reymonta 4, 30-059 Cracow, Poland}  }

\begin{document}
\maketitle

\interlinia

\begin{abstract}
\interlinia
Statistical theory of the dielectric susceptibility
of polar liquid crystals is proposed.
The molecules are calamitic or bent-core but the permanent dipole
moment is perpendicular to the molecule long axis.
The ordering of the phase is described by means of
the mean-field theory based on the Maier-Saupe approach.
The theory is used to calculate the temperature
dependence of the order parameters and the susceptibilities.
The phase diagram with four phases is obtained: isotropic,
uniaxial nematic, uniaxial ferroelectric, and biaxial ferroelectric.
Four critical points are predicted.
\end{abstract}

\section{Introduction}
\label{sec1}

Liquid crystals are always built from anisotropic molecules
\cite{[1993_de_Gennes]}.
For a long time calamitic or discotic molecules were considered
as appropriate molecules for the formation of liquid crystal phases.
But it appeared that bent-core molecules can lead to many
interesting phenomena related to polarity and chirality
\cite{[2006_Takezoe]}.
Phases and phases transitions that can take place in the bent-core 
systems were studied by Lubensky and Radzihowsky
\cite{[2002_Radzihovsky_Lubensky]}.
They showed that to completely characterize phases a third-rank
tensor order parameter is necessary in addition to the vector and
the nematic (second-rank) tensor order parameters.
Twelve different liquid phases were identified and many
symmerty-allowed transitions among them were analysed in detail.
The bent-core molecules were studied also by means of computer
simulations
\cite{[1999_Camp]}-\cite{[2007_Bates]}.

We would like to study a simple model that can capture
main features of bent-core molecules.
Let us consider a system of $N$ molecules contained in 
a volume $V$ at temperature $T$.
We assume that the potential energy of the interactions $V(R_1,R_2)$
depends only on the molecule orientations $R_1$ and $R_2$.
The orientation of a molecule is described by the three Euler
angles $R=(\phi,\theta,\psi)$ or by the three orthonormal
vectors $(\vec{l},\vec{m},\vec{n})$.
The potential energy of interactions has the form
\begin{equation}
\label{VR1R2}
V(R_1,R_2) = v_0
+ v_1 P_1(\vec{l}_1 \cdot \vec{l}_2)
+ v_2 P_2(\vec{n}_1 \cdot \vec{n}_2),
\end{equation}
where $P_j$ are the Legendre polynomials, the vector $\vec{n}$ 
determine the long molecule axis.
If $v_1=0$ then the Maier-Saupe theory is recovered, where the nematic
phase is present for $v_2 < 0$.
The $v_1$ term breaks $D_{\infty h}$ symmetry and it partly describes
bent-core molecules with $C_{2v}$ symmetry.
On the other hand, the $v_1$ term can be connected with the permanent
electric dipol $\vec{p}=p\vec{l}$ of a molecule because many bent-core
molecules have a transverse dipol which bisects the bend angle
\cite{[2007_Bates]}.

Our aim is to create the theory of the susceptibility which
is a tensor in the case of anisotropic phases
\begin{equation}
\label{chiab}
\epsilon_0 \chi_{\alpha\beta} = 
\frac{\partial P_{\alpha}}{\partial E_{\beta}},
\end{equation}
where $P$ and $E$ are the polarization and the electric field,
respectively.
In the first approximation, the polarization can be written as
\begin{equation}
P_{\alpha} = \frac{Np}{V} \langle \bar{l}_{\alpha} \rangle,
\end{equation}
where $p\langle \bar{l}_{\alpha} \rangle$ is the average value of the dipole
component in the $\alpha$-direction in the presence of the electric field.
Two different averages are involved here. 
The bar refers to the potential
energy of the dipole moment in the electric field and 
the brackets to the nematic potential.
The potential energy of the dipole $\vec{p}$ in the electric
field $\vec{E}$ is $\vec{p} \cdot \vec{E}$.
The linear approximation is applied to resolve the influence
of the electric field
\begin{equation}
\langle \bar{l}_{\alpha} \rangle = 
\langle l_{\alpha} (1+\beta p l_{\beta} E_{\beta}) \rangle =
\beta p \langle l_{\alpha} l_{\beta} \rangle E_{\beta}.
\end{equation}
Thus the susceptibility has the form
\begin{equation}
\epsilon_0 \chi_{\alpha\beta} = 
\frac{N p^2 \beta}{V} \langle l_{\alpha} l_{\beta} \rangle.
\end{equation}
The components of the susceptibility in the nematic phase
oriented along the z axis are
\begin{eqnarray}
\epsilon_0 \chi_{zz} &=& \frac{N p^2 (1-S)}{3V k_B T}, \\ 
\epsilon_0 \chi_{xx}&=& \frac{N p^2 (2+S)}{6V k_B T},
\end{eqnarray}
where $S=\langle P_2(n_z) \rangle$ is the order parameter.
More advanced calculations were given by Maier and Meier
\cite{[1961_Maier_Meier]}
who extended the Onsager theory of the susceptibility
to nematic liquid crystals. If we neglect the induced polarization
we can write their results as
\begin{eqnarray}
\epsilon_0 \chi_{zz} &=& 
\left(\frac{3 \tilde{\chi} + 3}{2 \tilde{\chi} +3} \right)
\frac{N p^2 (1-S)}{3V k_B T}, \\ 
\epsilon_0 \chi_{xx}&=& 
\left(\frac{3 \tilde{\chi} + 3}{2 \tilde{\chi} +3} \right)
\frac{N p^2 (2+S)}{6V k_B T},
\end{eqnarray}
where $\tilde{\chi}$ is the average susceptibility of the phase.

The organization of this paper is as follows:
In Sec. \ref{sec2} the mean-field theory of the phase ordering
is provided.
In Sec. \ref{sec3} the formulae for the susceptibility are derived
in the case of the nematic and the ferroelectric phase.
Section \ref{sec4} is devoted to some applications
of the presented theory.
Section \ref{sec5} contains a summary.
Appendix A provides the definitions and main properties
of the basic functions.

\section{Mean-field theory}
\label{sec2}

The potential energy of molecular interactions (\ref{VR1R2})
can be rewritten in the form
\begin{equation}
V(R_1,R_2) = v_0
+ v_1 E_{11}^{(1)}(R_2^{-1} R_1)
+ v_2 E_{00}^{(2)}(R_2^{-1} R_1),
\end{equation}
where $E_{\mu\nu}^{(j)}$ are the basic functions defined in Appendix~A
\cite{[2005_Kapanowski_Wietecha]}.
The basic functions will facilitate many future calculations.
The molecular orientation in the phase is described by the distribution 
function
\begin{equation}
\int \! {dR} f(R) = 1. 
\end{equation}
The mean of any function $A=A(R)$ we calculate as
\begin{equation}
\langle A \rangle  \equiv  \int \! {dR} f(R) A(R).
\end{equation}
The state of the system is described by a series of order parameters
$\langle E_{\mu\nu}^{(j)} \rangle$ but the most important
have $j=1$ or $j=2$.
The internal energy of the system is
\begin{equation}
U = \frac{N}{2} \int \! {dR_1} {dR_2} f(R_1) f(R_2) V(R_1,R_2),
\end{equation}
whereas the entropy of the system has the form
\begin{equation}
S = -k_B N \int \! {dR} f(R) \ln [f(R) C_N].
\end{equation}
The energy of permanent dipole moments 
in the electric field $\vec{E}$ is
\begin{equation}
U_E =  N \langle \vec{p}  \cdot \vec{E} \rangle.
\end{equation}
The total free energy of the system is the sum
\begin{equation}
F_{tot}=U+U_E-TS.
\end{equation}
In the mean-field approximation a potential energy is
\begin{equation}
W(R) = \sum_{j} \sum_{\mu\nu}
w_{\mu\nu}^{(j)} E_{\mu\nu}^{(j)}(R),
\end{equation}
and we find from the Boltzmann distribution that
\begin{equation}
f(R) = \exp [-\beta W(R)]/Z,
\end{equation}
where $Z$ is a normalization constant.
The consistency condition 
\begin{equation}
W(R_1) =  \int \! {dR_2} f(R_2) V(R_1,R_2)
-\vec{p}(R_1) \cdot \vec{E}
\end{equation}
leads to equations
\begin{eqnarray}
\label{wjmn1}
w_{\mu\nu}^{(1)} & = & 
v_1 \langle E_{\mu 1}^{(1)} \rangle \delta_{\nu 1}
-p \delta_{\nu 1} 
(E_x \delta_{1\mu} - E_y \delta_{-1\mu} + E_z \delta_{0\mu}),
\\
w_{\mu\nu}^{(2)} & = & 
v_2 \langle E_{\mu 0}^{(2)} \rangle \delta_{\nu 0}.
\label{wjmn2}
\end{eqnarray}
It is useful to introduce the dimensionless parameters
$S_{\mu\nu}^{(j)}= -\beta w_{\mu\nu}^{(j)}$ for $j>0$
\begin{equation}
\ln f(R) = \sum_j \sum_{\mu\nu} S_{\mu\nu}^{(j)} E_{\mu\nu}^{(j)}(R).
\end{equation}
$S_{00}^{(0)}$ is responsible for the normalization and it depends
on other $S_{\mu\nu}^{(j)}$ with $j>0$
\begin{equation} 
S_{00}^{(0)}= -\ln \left[\int {dR}
\exp \left( \sum_{j>0} \sum_{\mu\nu} 
S_{\mu\nu}^{(j)} E_{\mu\nu}^{(j)}(R) \right) \right] ,
\end{equation}
\begin{equation}
U_{\mu\nu}^{11} \equiv
\langle E_{\mu 1}^{(1)} E_{\nu 1}^{(1)} \rangle 
- \langle E_{\mu 1}^{(1)} \rangle \langle E_{\nu 1}^{(1)} \rangle ,
\end{equation}
\begin{equation}
U_{\mu\nu}^{12} \equiv
\langle E_{\mu 1}^{(1)} E_{\nu 0}^{(2)} \rangle 
- \langle E_{\mu 1}^{(1)} \rangle \langle E_{\nu 0}^{(2)} \rangle ,
\end{equation}
\begin{equation}
U_{\mu\nu}^{22} \equiv
\langle E_{\mu 0}^{(2)} E_{\nu 0}^{(2)} \rangle 
- \langle E_{\mu 0}^{(2)} \rangle \langle E_{\nu 0}^{(2)} \rangle ,
\end{equation}
\begin{equation}
\frac{\partial S_{00}^{(0)}}{\partial S_{\mu\nu}^{(j)}}
= -\langle E_{\mu\nu}^{(j)} \rangle,\ 
U_{\mu\nu}^{jk} = U_{\nu\mu}^{kj}.
\end{equation}
Now the equations (\ref{wjmn1})-(\ref{wjmn2}) have the form
\begin{eqnarray}
S_{\mu\nu}^{(1)}
+\beta v_1 \langle E_{\mu 1}^{(1)} \rangle \delta_{\nu 1} 
& = & \beta p \delta_{\nu 1} 
(E_x \delta_{1\mu} - E_y \delta_{-1\mu} + E_z \delta_{0\mu}),
\\
S_{\mu\nu}^{(2)} 
+\beta v_2 \langle E_{\mu 0}^{(2)} \rangle \delta_{\nu 0}
& = & 0.
\end{eqnarray}
The solution is orientationally stable only if the matrix
\begin{equation}
\left[
U_{\mu\nu}^{jk} + \sum_{l>0} \sum_{\rho} \beta v_l 
U_{\mu\rho}^{jl} U_{\nu\rho}^{kl}
\right]
\end{equation}
is positive definite. The isotropic phase is orientationally stable
if $\beta v_1 > -3$ 
and $\beta v_2 > -5$.

\section{Dielectric susceptibility}
\label{sec3}

The dielectric susceptibility tensor is defined by the Eq. 
(\ref{chiab}) and we calculate the orientational polarization as
\begin{equation}
P_{\alpha} = \langle l_{\alpha} \rangle N p/V.
\end{equation}
Note that the polarization depends on the electric field via the
distribution function.
The components of the susceptibility are
\begin{eqnarray}
\epsilon_0 \chi_{xx} & = & 
\frac{N p}{V} \left[
\sum_{\mu} U_{1\mu}^{11}
\frac{\partial S_{\mu 1}^{(1)}}{\partial E_x}
+\sum_{\mu} U_{1\mu}^{12}
\frac{\partial S_{\mu 0}^{(2)}}{\partial E_x}
\right], 
\\
\epsilon_0 \chi_{yy} & = & 
\frac{N p}{V} (-1) \left[
\sum_{\mu} U_{-1\mu}^{11}
\frac{\partial S_{\mu 1}^{(1)}}{\partial E_y}
+\sum_{\mu} U_{-1\mu}^{12}
\frac{\partial S_{\mu 0}^{(2)}}{\partial E_y}
\right], 
\\
\epsilon_0 \chi_{zz} & = & 
\frac{N p}{V} \left[
\sum_{\mu} U_{0\mu}^{11}
\frac{\partial S_{\mu 1}^{(1)}}{\partial E_z}
+\sum_{\mu} U_{0\mu}^{12}
\frac{\partial S_{\mu 0}^{(2)}}{\partial E_z}
\right],
\end{eqnarray}
where the derivatives are calculated from the equations
\begin{eqnarray}
\frac{\partial S_{\mu 1}^{(1)}}{\partial E_x}
+\beta v_1 \left[ 
\sum_{\rho} U_{\mu\rho}^{11} 
\frac{\partial S_{\rho 1}^{(1)}}{\partial E_x}
+\sum_{\rho} U_{\mu\rho}^{12} 
\frac{\partial S_{\rho 0}^{(2)}}{\partial E_x}
\right]
& = & \beta p  \delta_{1\mu}, 
\\
\frac{\partial S_{\mu 0}^{(2)}}{\partial E_x}
+\beta v_2 \left[ 
\sum_{\rho} U_{\mu\rho}^{21} 
\frac{\partial S_{\rho 1}^{(1)}}{\partial E_x}
+\sum_{\rho} U_{\mu\rho}^{22} 
\frac{\partial S_{\rho 0}^{(2)}}{\partial E_x}
\right]
& = & 0, 
\\
\frac{\partial S_{\mu 1}^{(1)}}{\partial E_y}
+\beta v_1 \left[ 
\sum_{\rho} U_{\mu\rho}^{11} 
\frac{\partial S_{\rho 1}^{(1)}}{\partial E_y}
+\sum_{\rho} U_{\mu\rho}^{12} 
\frac{\partial S_{\rho 0}^{(2)}}{\partial E_y}
\right]
& = & -\beta p  \delta_{-1\mu}, 
\\
\frac{\partial S_{\mu 0}^{(2)}}{\partial E_y}
+\beta v_2 \left[ 
\sum_{\rho} U_{\mu\rho}^{21} 
\frac{\partial S_{\rho 1}^{(1)}}{\partial E_y}
+\sum_{\rho} U_{\mu\rho}^{22} 
\frac{\partial S_{\rho 0}^{(2)}}{\partial E_y}
\right]
& = & 0, 
\\
\frac{\partial S_{\mu 1}^{(1)}}{\partial E_z}
+\beta v_1 \left[ 
\sum_{\rho} U_{\mu\rho}^{11} 
\frac{\partial S_{\rho 1}^{(1)}}{\partial E_z}
+\sum_{\rho} U_{\mu\rho}^{12} 
\frac{\partial S_{\rho 0}^{(2)}}{\partial E_z}
\right]
& = & \beta p  \delta_{0\mu}, 
\\
\frac{\partial S_{\mu 0}^{(2)}}{\partial E_z}
+\beta v_2 \left[ 
\sum_{\rho} U_{\mu\rho}^{21} 
\frac{\partial S_{\rho 1}^{(1)}}{\partial E_z}
+\sum_{\rho} U_{\mu\rho}^{22} 
\frac{\partial S_{\rho 0}^{(2)}}{\partial E_z}
\right]
& = & 0.
\end{eqnarray}
Now we are in the position to discuss the results for
possible phases.

\subsection{The isotropic phase}

For the zero field all order parameters are equal to zero.
For the nonzero field the phase has the symmetry 
$C_{\infty v}$ (the symmetry of the electric field).
\begin{equation}
\epsilon_0 \chi_{xx} 
= \epsilon_0 \chi_{yy} 
= \epsilon_0 \chi_{zz} 
= \frac{N p^2}{V(3 k_B T + v_1)}.
\end{equation}
For $v_1<0$ we get the Curie-Weiss law 
describing the divergence of $\chi$
when we approach the Curie temperature from above.
For $v_1>0$ the susceptibility is finite.

\subsection{The uniaxial nematic phase}

For the zero field the phase has the symmetry $D_{\infty h}$ and
the parameter $S_{00}^{(2)}$ is nonzero for the phase oriented
along the $z$ axis.
The nonzero order parameters are
$\langle E_{00}^{(j)} \rangle$ for $j$ even. 
For the parallel field the phase has the symmetry $C_{\infty v}$ 
whereas for the perpendicular field the phase has the symmetry $C_{2 v}$.
The expressions $U_{\mu\mu}^{jj}$ are nonzero.
\begin{eqnarray}
\epsilon_0 \chi_{xx} &=& 
\frac{N p^2 \beta U_{11}^{11}}{V (1+\beta v_1 U_{11}^{11})} ,
\\
\epsilon_0 \chi_{yy} &=& 
\frac{N p^2 \beta U_{-1,-1}^{11}}{V (1+\beta v_1 U_{-1,-1}^{11})} ,
\\
\epsilon_0 \chi_{zz} &=& 
\frac{N p^2 \beta U_{00}^{11}}{V (1+\beta v_1 U_{00}^{11})}.
\end{eqnarray}
Note that $\chi_{xx}=\chi_{yy}$.

\subsection{The uniaxial ferroelectric phase}

For the zero field and for the parallel field the phase has 
the symmetry $C_{\infty v}$.
The phase is oriented along the $x$ axis.
We have nonzero $S_{11}^{(1)}$ and
$S_{20}^{(2)} = -\sqrt{3} S_{00}^{(2)}$.
The nonzero order parameters are:
$\langle E_{11}^{(1)} \rangle$,
$\langle E_{20}^{(2)} \rangle = -\sqrt{3} \langle E_{00}^{(2)} \rangle$,
$\langle E_{02}^{(2)} \rangle$,
$\langle E_{22}^{(2)} \rangle$.
In the ideally ordered phase the values of the order parameters are:
$\langle E_{11}^{(1)} \rangle = 1$,
$\langle E_{00}^{(2)} \rangle = 1/4$,
$\langle E_{02}^{(2)} \rangle = \langle E_{20}^{(2)} \rangle = -\sqrt{3}/4$,
$\langle E_{22}^{(2)} \rangle = 3/4$.
The most important nonzero elements $U_{\mu\nu}^{jk}$ are:
$U_{11}^{11}$, 
$U_{00}^{11}$, 
$U_{-1,-1}^{11}$, 
$U_{12}^{12}$, 
$U_{10}^{12}$, 
$U_{01}^{12}$, 
$U_{-1,-2}^{12}$, 
$U_{22}^{22}$, 
$U_{20}^{22}$, 
$U_{00}^{22}$, 
$U_{11}^{22}$, 
$U_{-1,-1}^{22}$, 
$U_{-2,-2}^{22}$.
\begin{equation}
\label{chixx-ferro}
\epsilon_0 \chi_{xx} = 
\left(   \frac{N p^2 \beta}{V}  
\right)
\frac{U_{11}^{11} A_1 - U_{12}^{12} A_2 + U_{10}^{12} A_3
}{
(1+\beta v_1 U_{11}^{11}) A_1
- \beta v_1 U_{12}^{12} A_2
+ \beta v_1 U_{10}^{12} A_3
},
\end{equation}
where
\begin{eqnarray}
A_1 &=& 
(1+\beta v_2 U_{22}^{22}) (1+\beta v_2 U_{00}^{22})
-(\beta v_2 U_{02}^{22}) (\beta v_2 U_{20}^{22}),
\\
A_2 &=& 
(\beta v_2 U_{21}^{21}) (1+\beta v_2 U_{00}^{22})
-(\beta v_2 U_{01}^{21}) (\beta v_2 U_{20}^{22}),
\\
A_3 &=& 
(\beta v_2 U_{21}^{21}) (\beta v_2 U_{02}^{22})
-(\beta v_2 U_{01}^{21}) (1+\beta v_2 U_{22}^{22}).
\end{eqnarray}

\subsection{The biaxial ferroelectric phase}

For the zero field and for the parallel field the phase has 
the symmetry $C_{2 v}$.
The phase is oriented along the $x$ axis.
We have independent nonzero $S_{11}^{(1)}$,
$S_{20}^{(2)}$, and  $S_{00}^{(2)}$.
The most important order parameters are:
$\langle E_{11}^{(1)} \rangle$,
$\langle E_{00}^{(2)} \rangle$,
$\langle E_{02}^{(2)} \rangle$,
$\langle E_{20}^{(2)} \rangle$,
$\langle E_{22}^{(2)} \rangle$, and
$\langle E_{-1,-1}^{(2)} \rangle$.
The susceptibility is described by Eq. (\ref{chixx-ferro}).

\section{Results}
\label{sec4}

In this section we carry out calculations for different
physical systems of polar molecules described 
by the considered model.
The phase diagram of the model is shown in Fig. 1.
Four phases are present: isotropic, uniaxial nematic,
uniaxial ferroelectric, and biaxial ferroelectric.
Dashed half-lines in the picture denote different
physical systems with the fixed parameters $v_1$ and $v_2$.
Four critical points from $C_1$ to $C_4$ are predicted.
$C_1 = (-3.0, -3.3)$ and $C_2 = (-3.4, -3.8)$ are tricritical
points where transitions changes from second-order to first-order.
$C_3 = (-3.0, -4.1)$ and $C_4 = (-2.5, -4.54)$ are points
where three phases coexist in equilibrium.

Let us start from the systems known from the previous studies
\cite{[2004_Kapanowski_Wietecha]}.
For the case of $v_2=0$ [half line (a) in Fig. 1]
we get the simple system with the second-order transition from
the isotropic to the uniaxial ferroelectric phase at
$T_C=-v_1/3k_B$.
For the case of $v_1=0$ [half line (d) in Fig. 1]
the interactions are uniaxial (small dipols)
and there is the first-order
transition from the isotropic to the uniaxial nematic phase.
The temperature dependence of the inversed susceptibility 
is presented in Fig. 2.
In all pictures, $T$ denotes the dimensionless temperature.
$T=1$ corresponds to the transition from the isotropic
to the nematic or ferroelectric phase.
The susceptibilities are expressed in
$N p^2/(V \epsilon_0 |v_1|)$.
On decreasing the temperature the susceptibility splits into
$\chi_{zz} < \chi_{xx}$. $\chi_{xx}$
runs to the infinity whereas $\chi_{zz}$ remains finite.
This is typical for the uniaxial molecules with the dipol
moment perpendicular to the symmetry axis.

In the system described by half line (b) in Fig. 1
there are the second-order transitions from the isotropic phase
to the uniaxial ferroelectric phase ($T=1$) and next to the biaxial
ferroelectric phase ($T=0.81$).
The temperature dependence of the order parameters and the inversed
susceptibility is presented in Fig. 3 and 4, respectively.
We note that the susceptibility reveals the transition from the
uniaxial ferroelectric to the biaxial ferroelectric phase.

A very interesting situation takes place in the system
described by half line (c) in Fig. 1.
The temperature dependence of the order parameters and the inversed
susceptibility is presented in Fig. 5 and 6, respectively.
On decreasing the temperature we meet the first-order transition
from the isotropic to the uniaxial nematic phase ($T=1$)
and the second-order transition to the biaxial ferroelectric phase ($T=0.96$).
In the uniaxial nematic phase we have $\chi_{xx} > \chi_{yy}=\chi_{zz}$.

In the case of the system described by half line (e) in Fig. 1
two phases are present: isotropic and uniaxial nematic with
the first-order transition between them.
The order parameter $\langle E_{00}^{(2)} \rangle$ is the most important
and the susceptibility is finite.

\section{Summary}
\label{sec5}

In this paper, we presented the statistical theory 
of the dielectric susceptibility of polar liquid crystals.
The molecules were calamitic or bent-core with the permanent
perpendicular dipole moment.
We focused on spatially homogeneous (liquid) phases
because the understanding and classifying of those phases
is a prerequisite to study of more ordered phases,
which in addition break translational symmetry.

Liquid phases of bent-core molecules from the considered model
are listed in Table 1. Phase transition types obtained in the present
paper can be summarized as follows:
\begin{enumerate}
\item
$I \rightarrow N$ transition: first-order;
\item
$I \rightarrow V$ transition: second-order or weakly first-order
(second-order in \cite{[2002_Radzihovsky_Lubensky]});
\item
$I \rightarrow V+2$ transition: first-order
(not present in \cite{[2002_Radzihovsky_Lubensky]});
\item
$N \rightarrow V+2$ transition: second-order;
\item
$V \rightarrow V+2$ transition: second-order or weakly first-order
(second-order in \cite{[2002_Radzihovsky_Lubensky]}).
\end{enumerate}
Every transition has unique features which can be observed from the 
temperature dependence of the order parameters or susceptibilities.
Generally, the classification by Lubensky and Radzihovsky 
\cite{[2002_Radzihovsky_Lubensky]}
is completed. 
The results for the transitions $I \rightarrow V$, $I \rightarrow V+2$,
and $V \rightarrow V+2$
was obtained also by Mettout {\em et al}
\cite{[2002_Mettout]}
in a theory with two vectors representing the six-dimentional
order parameter associated with the transition from the isotropic
liquid to a polar nematic phase.

Polar nematic phases of thermotropic liquid crystals may have
interesting technological applications, and that is why
a number of theoretical studies have focused on the possibility
of realizing such phases
\cite{[1988_Palffy]}-\cite{[2000_Brand]}.
One may expect for such phases, a rich variety of unusual textures
and defects.
The existence of biaxial and uniaxial ferroelectric phases
requires further experimental confirmation.

An extended version of the considered model can be obtained
from the Straley model of biaxial nematics
\cite{[1974_Straley]}
by the inclusion of our $P_1(\vec{l}_1 \cdot \vec{l}_2)$ term
or the $P_1(\vec{n}_1 \cdot \vec{n}_2)$ term.
Then the biaxial nematic $N+2$ phase with $D_{2h}$ symmetry
would be present and new transition types would be possible:
$I \rightarrow N+2$,
$N \rightarrow N+2$, or
$N+2 \rightarrow V+2$.

\section*{Appendix A}
\label{app1}

Below we list the properties of the functions $E_{\mu\nu}^{(j)}$.
The functions can be used to describe any physical quantity
which depends on the three Euler angles.

\begin{enumerate}
\item 
The definition is 
\begin{eqnarray}
& E_{\mu\nu}^{(j)}(R) = 
\left( 
{\frac{1}{\sqrt{2}}} 
\right)^{2+\delta_{0\mu}+\delta_{0\nu}}
{\frac{1}{2}}
[(1+i)+(1-i) \mbox{sign}(\mu) \mbox{sign}(\nu)]
& \nonumber\\
& \times [ D_{\mu\nu}^{(j)}(R)
+ \mbox{sign}(\mu) \mbox{sign}(\nu) (-1)^{\mu+\nu} 
D_{-\mu,-\nu}^{(j)}(R) 
& \nonumber\\
& \mbox{} + \mbox{sign}(\nu) (-1)^{\nu} D_{\mu,-\nu}^{(j)}(R)
+ \mbox{sign}(\mu) (-1)^{\mu} D_{-\mu,\nu}^{(j)}(R) ], &
\end{eqnarray}
where $R=(\phi,\theta,\psi)$ (the three Euler angles),
$j$ is a non-negative integer, 
$\mu$ and $\nu$ are integers.
Functions $D_{\mu\nu}^{(j)}$ are 
standard rotation matrix elements \cite{[1957_Edmonds]} and
\begin{equation}
\mbox{sign}(x)=
\left\{
\begin{array}{rl}
 1 & \mbox{for}\ x \ge 0 \\ 
 -1 & \mbox{for}\ x < 0. 
\end{array}
\right.
\end{equation}
Note that
\begin{equation} 
\mbox{sign}(-x) = - \mbox{sign}(x) + 2\delta_{0x}. 
\end{equation}
\item 
The functions $E_{\mu\nu}^{(j)}$ are real.
\item
The functions satisfy the orthogonality relations
\begin{equation}
\int {dR} E_{\mu\nu}^{(j)}(R) E_{\rho\sigma}^{(k)}(R)=
\delta_{jk} \delta_{\mu\rho} \delta_{\nu\sigma}
8\pi^{2}/(2j+1).
\end{equation}
\item
Let us assume that the three Euler angles $R=(\phi,\theta,\psi)$
determine the orientation of the three unit orthogonal vectors
$(\vec{l},\vec{m},\vec{n})$. 
The functions $E_{\mu\nu}^{(j)}$ can be expressed in terms 
of the vector coordinates.
\begin{equation}
E_{00}^{(1)} (R) = n_z,
\end{equation}
\begin{equation}
E_{01}^{(1)} (R) = l_z,
\end{equation}
\begin{equation}
E_{10}^{(1)} (R) = n_x,
\end{equation}
\begin{equation}
E_{11}^{(1)} (R) = l_x,
\end{equation}
\begin{equation}
E_{00}^{(2)} (R) = {\frac {1}{2}} (-1+3 n_{z}^2),
\end{equation}
\begin{equation}
E_{02}^{(2)} (R) = {\frac {\sqrt{3}}{2}} (-1+ n_{z}^2 + 2 l_{z}^2),
\end{equation}
\begin{equation}
E_{20}^{(2)} (R) = {\frac {\sqrt{3}}{2}} (-1+ n_{z}^2 + 2 n_{x}^2),
\end{equation}
\begin{equation}
E_{22}^{(2)} (R) = {\frac {1}{2}} (-3 + n_{z}^2 + 2 l_{z}^2 + 2 n_{x}^2 + 4 l_{x}^2).
\end{equation}
\end{enumerate}

\begin{figure}
\begin{center}
\includegraphics{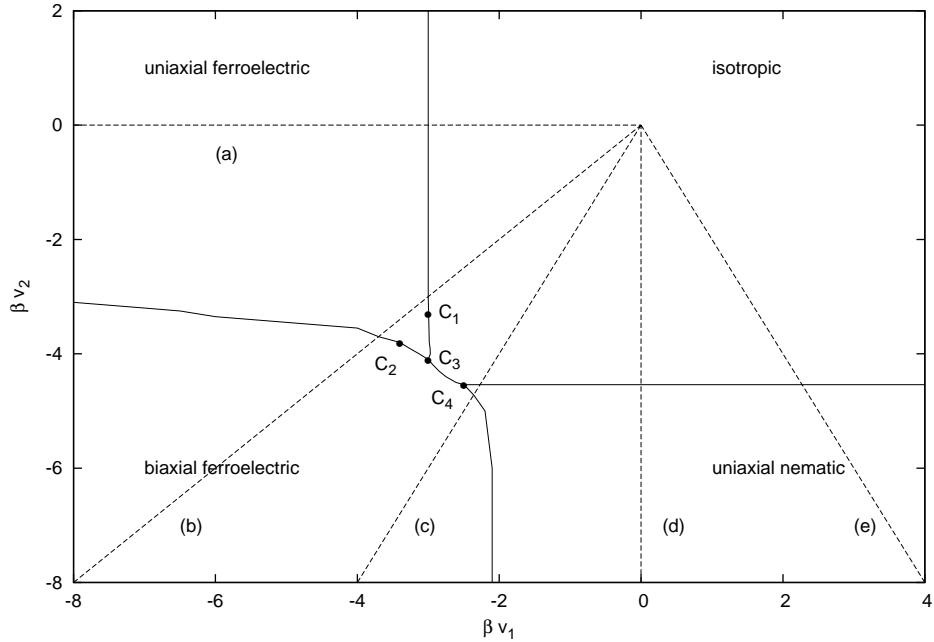}
\end{center}
\caption[Phase diagram of the model.]{
\label{fig1}
\interlinia
Phase diagram of the model considered in the paper, where
$v_1$ and $v_2$ are the parameters of molecular interactions,
$1/\beta = k_B T$.
Four phases are present: isotropic, uniaxial nematic,
uniaxial ferroelectric, and biaxial ferroelectric.
Four critical points from $C_1$ to $C_4$ are predicted.
Dashed half-lines denote different physical systems:
(a) $v_2=0$,
(b) $v_2=v_1$,
(c) $v_2=2 v_1$,
(d) $v_1=0$, and
(e) $v_2=-2 v_1$.
For a given physical system on decreasing the temperature
we are moving from the center (0,0) to the edge of the figure.}
\end{figure}

\begin{figure}
\begin{center}
\includegraphics{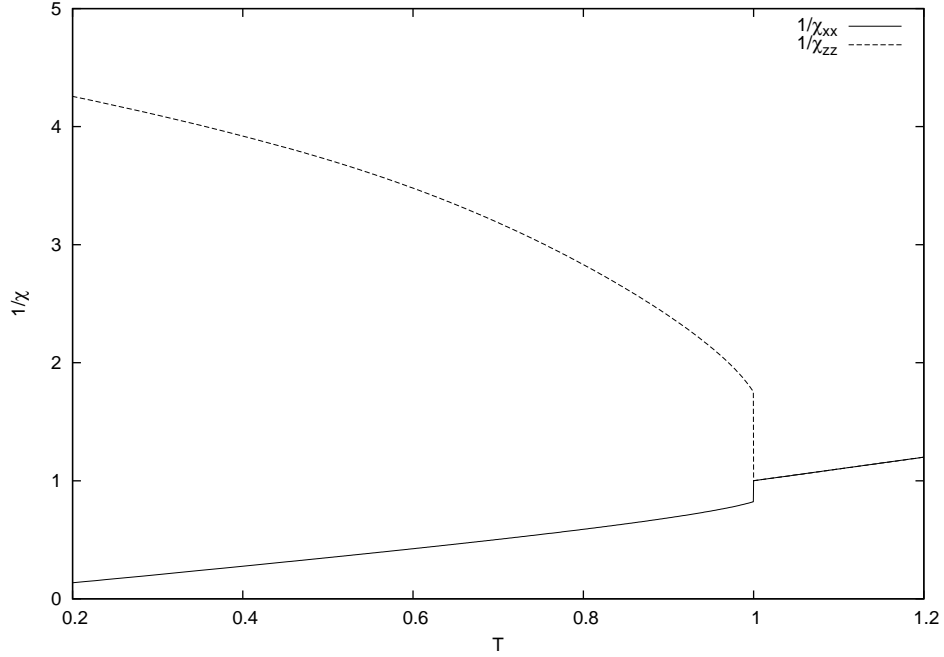}
\end{center}
\caption[Temperature dependence of the inversed susceptibility 
for \mbox{$v_1=0$}.]{
\label{fig2}
\interlinia
Temperature dependence of the inversed susceptibility for $v_1=0$
[half line (d) in Fig. 1].
On decreasing the temperature the susceptibility splits into
$\chi_{zz} < \chi_{xx}$.}
\end{figure}

\begin{figure}
\begin{center}
\includegraphics{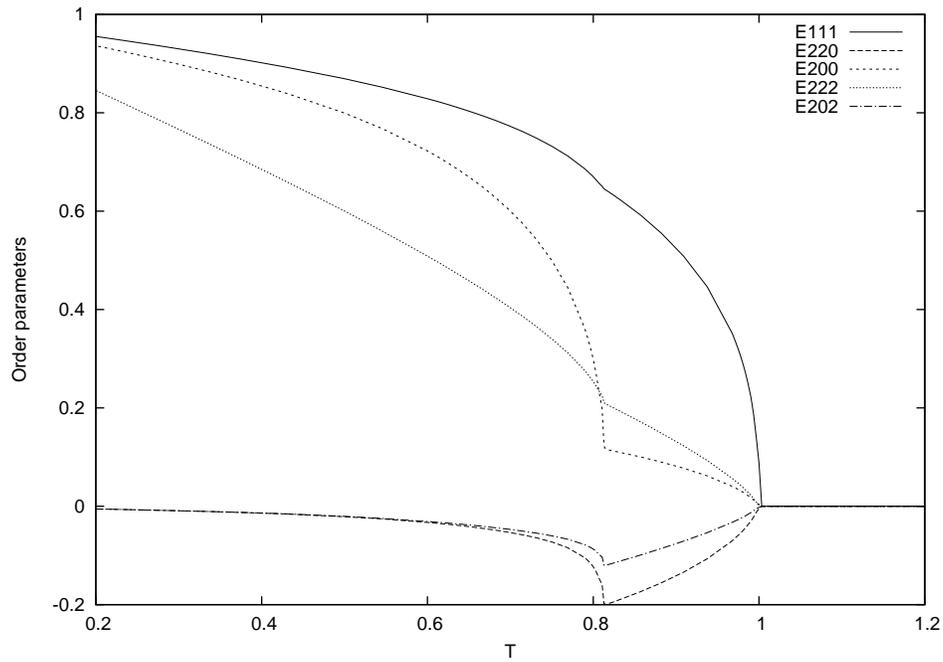}
\end{center}
\caption[Temperature dependence of the order parameters 
for \mbox{$v_1=v_2$}.]{
\label{fig3}
\interlinia
Temperature dependence of the order parameters for $v_1=v_2$
[half line (b) in Fig. 1].
There are the second-order transitions from the isotropic phase
to the uniaxial ferroelectric phase ($T=1$) and next to the biaxial
ferroelectric phase ($T=0.81$).}
\end{figure}

\begin{figure}
\begin{center}
\includegraphics{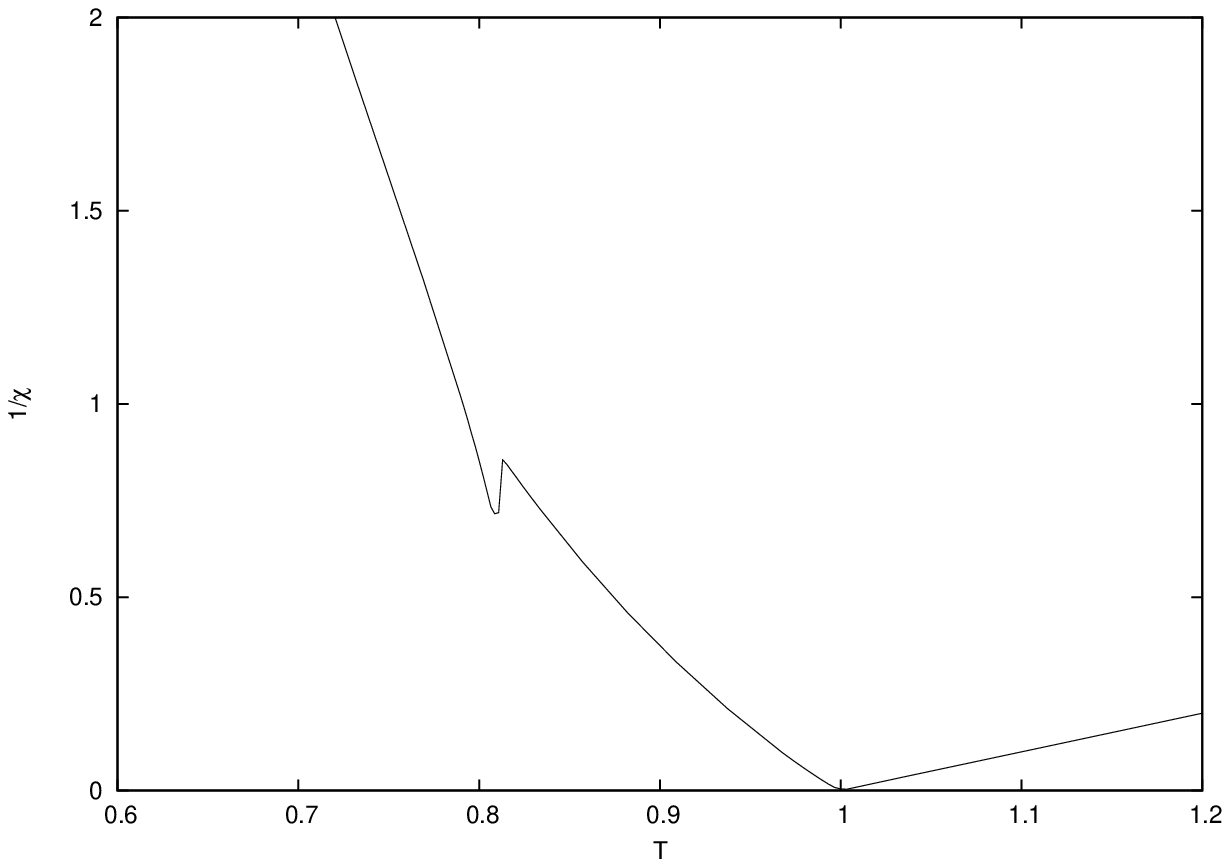}
\end{center}
\caption[Temperature dependence of the inversed susceptibility 
for \mbox{$v_1=v_2$}.]{
\label{fig4}
\interlinia
Temperature dependence of the inversed susceptibility for $v_1=v_2$
[half line (b) in Fig. 1].}
\end{figure}

\begin{figure}
\begin{center}
\includegraphics{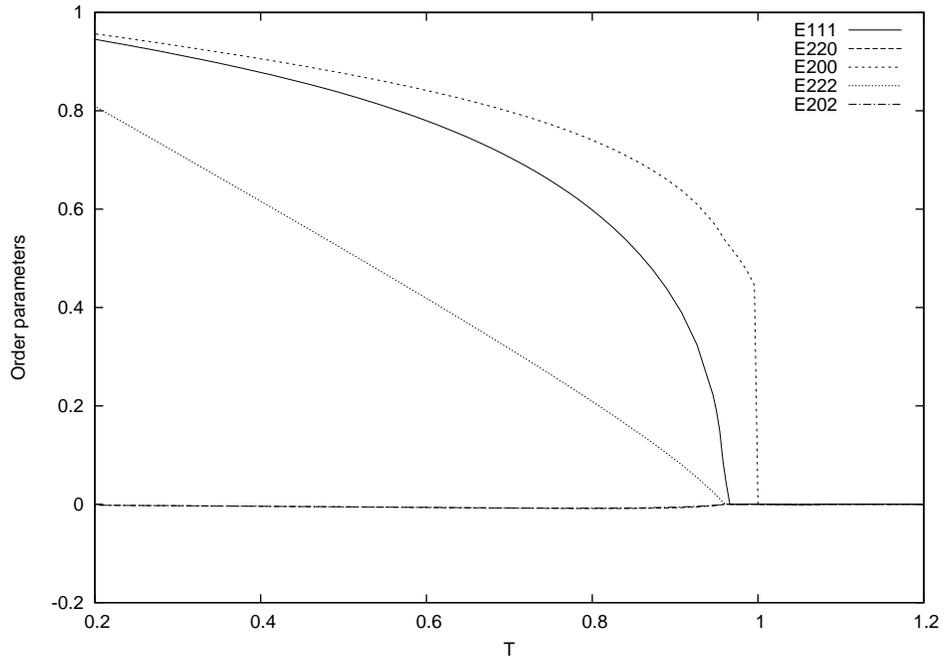}
\end{center}
\caption[Temperature dependence of the order parameters for 
\mbox{$v_2 = 2 v_1$}.]{
\label{fig5}
\interlinia
Temperature dependence of the order parameters for $v_2 = 2 v_1$
[half line (c) in Fig. 1].
There is the first-order transition from the isotropic phase
to the uniaxial nematic phase ($T=1$) and the second-order transition
from the uniaxial nematic phase to the biaxial
ferroelectric phase ($T=0.96$).}
\end{figure}

\begin{figure}
\begin{center}
\includegraphics{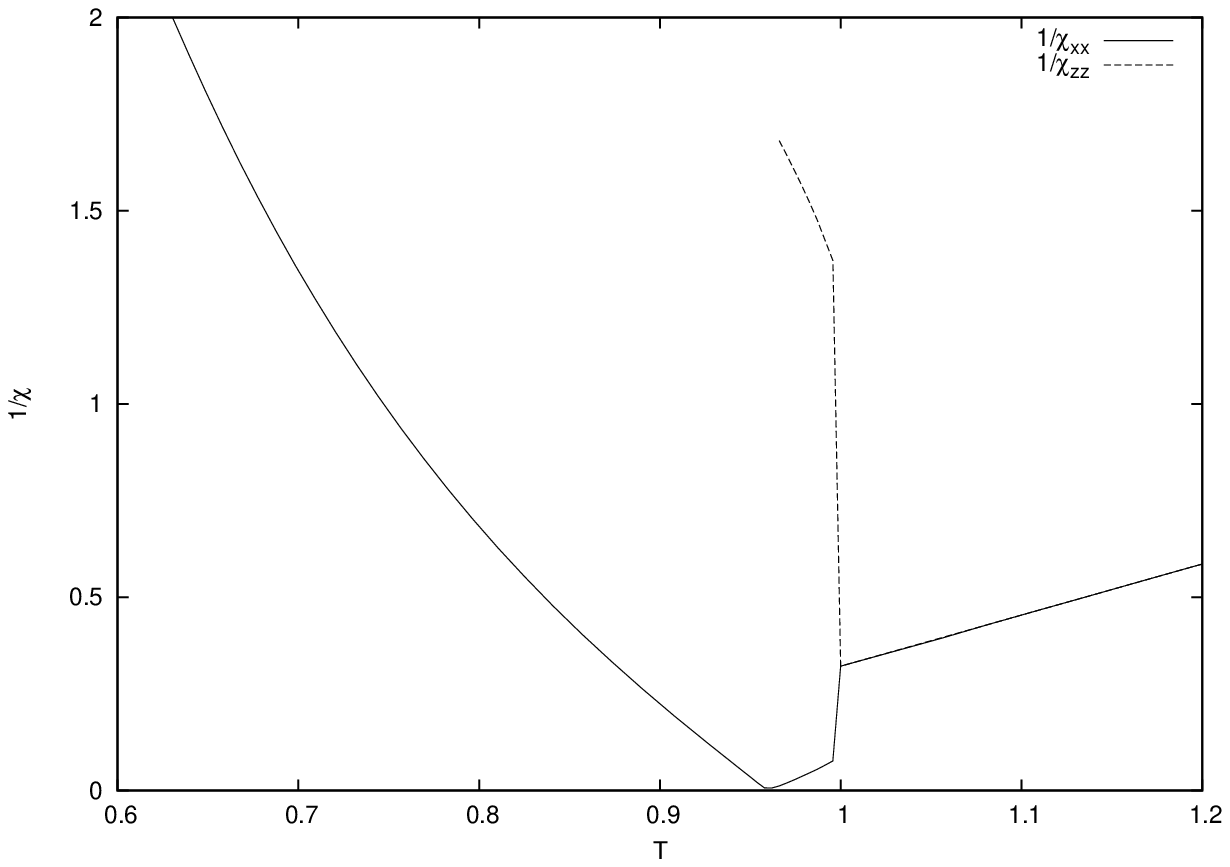}
\end{center}
\caption[Temperature dependence of the inversed susceptibility 
for \mbox{$v_2 = 2 v_2$}.]{
\label{fig6}
\interlinia
Temperature dependence of the inversed susceptibility for $v_2 = 2 v_1$
[half line (c) in Fig. 1].}
\end{figure}

\begin{table}
\caption{
\label{tab1}
\interlinia
Liquid phases of banana-shaped molecules, their symmetries 
in Schoenflies notation, and nonzero variables in the considered
model. Four phases are present: isotropic ($I$),
uniaxial nematic ($N$),
uniaxial ferroelectric ($V$), and
biaxial ferroelectric ($V+2$).}
\begin{center}
\begin{tabular}{ccc}
\hline\hline
Phase & Symmetry & Variables
\\ \hline
$I$   & $O(3)$        & - \\
$N$   & $D_{\infty h}$ & $S_{00}^{2}$ \\
$V$   & $C_{\infty v}$ & $S_{11}^{1}$, $S_{20}^{2}=-\sqrt{3} S_{00}^{2}$ \\
$V+2$ & $C_{2v}$       & $S_{11}^{1}$, $S_{00}^{2}$, $S_{20}^{2}$ \\
\hline\hline
\end{tabular}
\end{center}
\end{table}

\end{document}